\documentclass[preprint,showpacs,preprintnumbers,amsmath,amssymb]{revtex4}
\usepackage{amsmath}

\usepackage{graphicx}
\usepackage{pdfpages}
\usepackage{dcolumn}
\usepackage{bm}
\usepackage{amsmath} 
\usepackage{amsfonts}
\usepackage{amssymb}
\usepackage{url}
\usepackage{microtype}
\usepackage{graphicx}%

\newcommand{\qed}{\nobreak \ifvmode \relax \else
      \ifdim\lastskip<1.5em \hskip-\lastskip
      \hskip1.5em plus0em minus0.5em \fi \nobreak
      \vrule height0.75em width0.5em depth0.25em\fi}
\newcommand{\norm}[1]{\left\lVert#1\right\rVert}

\begin{document}

\preprint{}
\title{Density Matrix Diagonal-Block Lovas-Andai-type singular-value ratios for qubit-qudit separability/PPT probability analyses}
\author{Paul B. Slater}
 \email{slater@kitp.ucsb.edu}
\affiliation{%
Kavli Institute for Theoretical Physics, University of California, Santa Barbara, CA 93106-4030\\
}
\date{\today}
            
\begin{abstract}
An important variable in the 2017 analysis of Lovas and Andai, formally establishing the Hilbert-Schmidt separability probability conjectured by Slater of $\frac{29}{64}$ for the 9-dimensional convex set of two-rebit density matrices, was the ratio ($\varepsilon =\frac{\sigma_2}{\sigma_1}$) of the  two singular values ($\sigma_1 \geq \sigma_2 \geq 0$) of $D_2^{\frac{1}{2}} D_1^{-\frac{1}{2}}$. There, $D_1$ and $D_2$ were the  diagonal $2  \times 2$ blocks of a $4 \times 4$ two-rebit density matrix $\rho$. 
Working within the Lovas-Andai "separability function" ($\tilde{\chi}_d(\varepsilon)$) framework, Slater was able to verify further conjectures of Hilbert-Schmidt separability probabilities of $\frac{8}{33}$ and $\frac{26}{323}$ for the 15-dimensional and 26-dimensional convex sets of two-qubit and two-quater[nionic]-bit density matrices. Here, we investigate the behavior of the three singular value ratios of $V=D_2^{\frac{1}{2}} D_1^{-\frac{1}{2}}$, where now $D_1$ and $D_2$ are the $3 \times 3$ diagonal blocks of  $6 \times 6$ rebit-retrit and qubit-qutrit  density matrices randomly generated with respect to Hilbert-Schmidt measure. Further, we initiate a parallel study employing $8 \times 8$ density matrices. The motivation for this analysis is the conjectured relevance of these singular values in suitably extending  $\tilde{\chi}_d(\varepsilon)$ to higher dimensional systems--an issue we also approach using certain novel numeric  means. Section 3.3 of the 2017 A. Lovas doctoral dissertation (written in Hungarian) appears germane to such an investigation.
\end{abstract}


\maketitle
\section{Introduction}
Lovas and Andai formally demonstrated--as previously conjectured by Slater, on the basis of a diversity of prior analyses \cite{slater2013concise}--that the Hilbert-Schmidt (HS) separability probability of the 9-dimensional convex set of two-rebit systems is $\frac{29}{64} =\frac{29}{2^6} \approx 0.453125$ \cite{lovas2017invariance}. Their specific analytical structure  was then extended and applied by Slater to support--though not employing a similar lemma-theorem format--the further prior conjectures that the HS separability probabilities for the 15-dimensional two-qubit and 27-dimensional two-quater[nionic]-bit systems are $\frac{8}{33}=\frac{2^3}{3 \cdot 11} \approx 0.242424$ and $\frac{26}{323}=\frac{2 \cdot 13}{17 \cdot 19} \approx 0.0804954$, respectively \cite{slater2017master}. As earlier found \cite[eqs. (1)-(3)]{slater2013concise}, these three rational separability probability results are the cases $\alpha= \frac{1}{2},1,2$ of
\begin{equation} \label{Hou1}
\mathcal{P}_{sep/PPT}(\alpha) =\Sigma_{i=0}^\infty f(\alpha+i),
\end{equation}
where
\begin{equation} \label{Hou2}
f(\alpha) = \mathcal{P}_{sep/PPT}(\alpha)-\mathcal{P}_{sep/PPT}(\alpha +1) = \frac{ q(\alpha) 2^{-4 \alpha -6} \Gamma{(3 \alpha +\frac{5}{2})} \Gamma{(5 \alpha +2})}{3 \Gamma{(\alpha +1)} \Gamma{(2 \alpha +3)} 
\Gamma{(5 \alpha +\frac{13}{2})}},
\end{equation}
and
\begin{equation} \label{Hou3}
q(\alpha) = 185000 \alpha ^5+779750 \alpha ^4+1289125 \alpha ^3+1042015 \alpha ^2+410694 \alpha +63000 = 
\end{equation}
\begin{displaymath}
\alpha  \bigg(5 \alpha  \Big(25 \alpha  \big(2 \alpha  (740 \alpha
   +3119)+10313\big)+208403\Big)+410694\bigg)+63000.
\end{displaymath}
Here, $\alpha = \frac{d}{2}$, where $d$ is the Dyson index of random matrix theory.

Equivalently, one can  employ \cite[p. 26]{slater2013concise}
\begin{equation} \label{InducedMeasureCase}
\mathcal{P}_{sep/PPT}(0,d)= 2 Q(0,d)= 1-    
\frac{\sqrt{\pi } 2^{-\frac{9 d}{2}-\frac{5}{2}} \Gamma \left(\frac{3 (d+1)}{2}\right)
   \Gamma \left(\frac{5 d}{4}+\frac{19}{8}\right) \Gamma (2 d+2) \Gamma \left(\frac{5
   d}{2}+2\right)}{\Gamma (d)} \times
\end{equation}
\begin{displaymath}
\, _6\tilde{F}_5\left(1,d+\frac{3}{2},\frac{5 d}{4}+1,\frac{1}{4} (5 d+6),\frac{5
   d}{4}+\frac{19}{8},\frac{3 (d+1)}{2};\frac{d+4}{2},\frac{5
   d}{4}+\frac{11}{8},\frac{1}{4} (5 d+7),\frac{1}{4} (5 d+9),2 (d+1);1\right)
\end{displaymath}
(the tilde indicating regularization of the hypergeometric function).

Also for even
$d$, Dunkl has obtained \cite[App. D]{slater2017master}
\[
\mathcal{P}_{sep/PPT}\left(  d\right)  =3456^{d}\frac{\left(  \frac{1}{2}\right)  _{d/2}%
^{3}\left(  \frac{7}{6}\right)  _{d/2}^{2}\left(  \frac{5}{6}\right)
_{d/2}^{2}\left(  2d\right)  !}{\left(  \frac{d}{2}\right)  !\left(  3\right)
_{5d}}\sum_{i\geq0,j\geq0}^{i+j\leq d/2}\frac{\left(  -\frac{d}{2}\right)
_{i+j}\left(  \frac{d}{2}\right)  _{j}\left(  d\right)  _{j}\left(
2+3d\right)  _{i}\left(  1+d\right)  _{i}}{\left(  2+\frac{5d}{2}\right)
_{i+j}\left(  1+\frac{d}{2}\right)  _{j}i!j!\left(  -2d\right)  _{i}}.
\]
C. Koutschan, using his HolonomicFunctions program \cite{koutschan2013creative}, established that these three separability probability 
formulas satisfy the same order-4 recurrence, essentially demonstrating their equivalence.

Here, we begin by investigating whether the lines of analysis recently pursued by Lovas and Andai and 
Slater in \cite{lovas2017invariance,slater2017master} can be "lifted" in a productive manner to the 20-dimensional rebit-retrit and 35-dimensional qubit-qutrit settings. (Prior  conjectures for the associated HS separability probabilities are $\frac{860}{6561}=\frac{2^2 \cdot 5 \cdot 43}{3^8} \approx 0.1310775$ and $\frac{27}{1000} =\frac{3^3}{2^3 \cdot 5^3}=0.027$, respectively  \cite{slater2021rational,slater2019numerical,slater2019extensions}.)

Central to the Lovas-Andai analysis--and motivating our present research--was the singular-value ratio \cite[Lemma 5, p. 7]{lovas2017invariance}
\begin{equation}
\varepsilon \equiv \sigma(V) =  \exp \left({-\cosh^{-1}\left(
		\frac{||V||^2_{HS}}{2 \det (V)}
		\right)}\right)= \exp \left({-\cosh^{-1}\left(
		\frac{1}{2}
		\sqrt{\frac{\det (D_1)}{\det (D_2)}}
		\mbox{Tr} \left(D_2 D_1^{-1}\right)
		\right)}\right),  
\end{equation}
of the $2 \times 2$ matrix  $V=D_2^{\frac{1}{2}} D_1^{-\frac{1}{2}}$, where $D_1$ and $D_2$ are the two diagonal blocks of a $4 \times 4$ density matrix $\rho$.
(Also, interestingly. the quite different function of these $2 \times 2$ matrices, $D=D_1+D_2$, plays an important role too. As conjectured by Milz and Strunz \cite{milz2014volumes}, and formally demonstrated by Lovas and Andai \cite[Cor. 2]{lovas2017invariance}, the separability probability of two-qubit--and qubit-qutrit states depends only on the Bloch radius ($r$) of $D$ and is constant in $r$.)
\section{Analyses}
To begin our study here, we analyze the  {\it three} singular-value ratios, where now  $V=D_2^{\frac{1}{2}} D_1^{-\frac{1}{2}} $, and $D_1, D_2$ are the $3 \times 3$ diagonal blocks of a $6 \times 6$ density matrix   $\rho$. Let us note that in Appendix B, "Rebit-retrit and qubit-qutrit analyses" of \cite{slater2017master}--motivated by a "Dyson-index-based ansatz" \cite[eq. 8)]{slater2017master}--we exhibited three plots--based on billions of randomly generated density matrices--of ratios of the squares of rebit-retrit separability probabilities to qubit-qutrit separability probabilities. The first plot employed $\sqrt{\frac{\rho_ {11} \rho_{66}}{\rho_ {33} \rho_ {44}}}$ as the independent variable, another plot
employed $\tau_1=\sqrt{\frac{\rho_ {11} \rho_{55}}{\rho_ {22} \rho_ {44}}}$  and $\tau_2 =\sqrt{\frac{\rho_ {22} d _{66}}{\rho_ {33} d _{55}}}$ as independent variables, and the third, the ratios of the second largest to the largest singular value, and the third largest to the second largest. 

\subsection{Joint distribution of singular-value ratios between separable and entangled states}
We generated density matrices randomly with respect to Hilbert-Schmidt measure \cite{al2010random}--and assigned them to different sets depending upon whether they corresponded to separable or entangled states.
When each set had twenty thousand members, we terminated the random generation process. 
\subsubsection{Rebit-retrit case}
\begin{figure}
    \centering
    \includegraphics{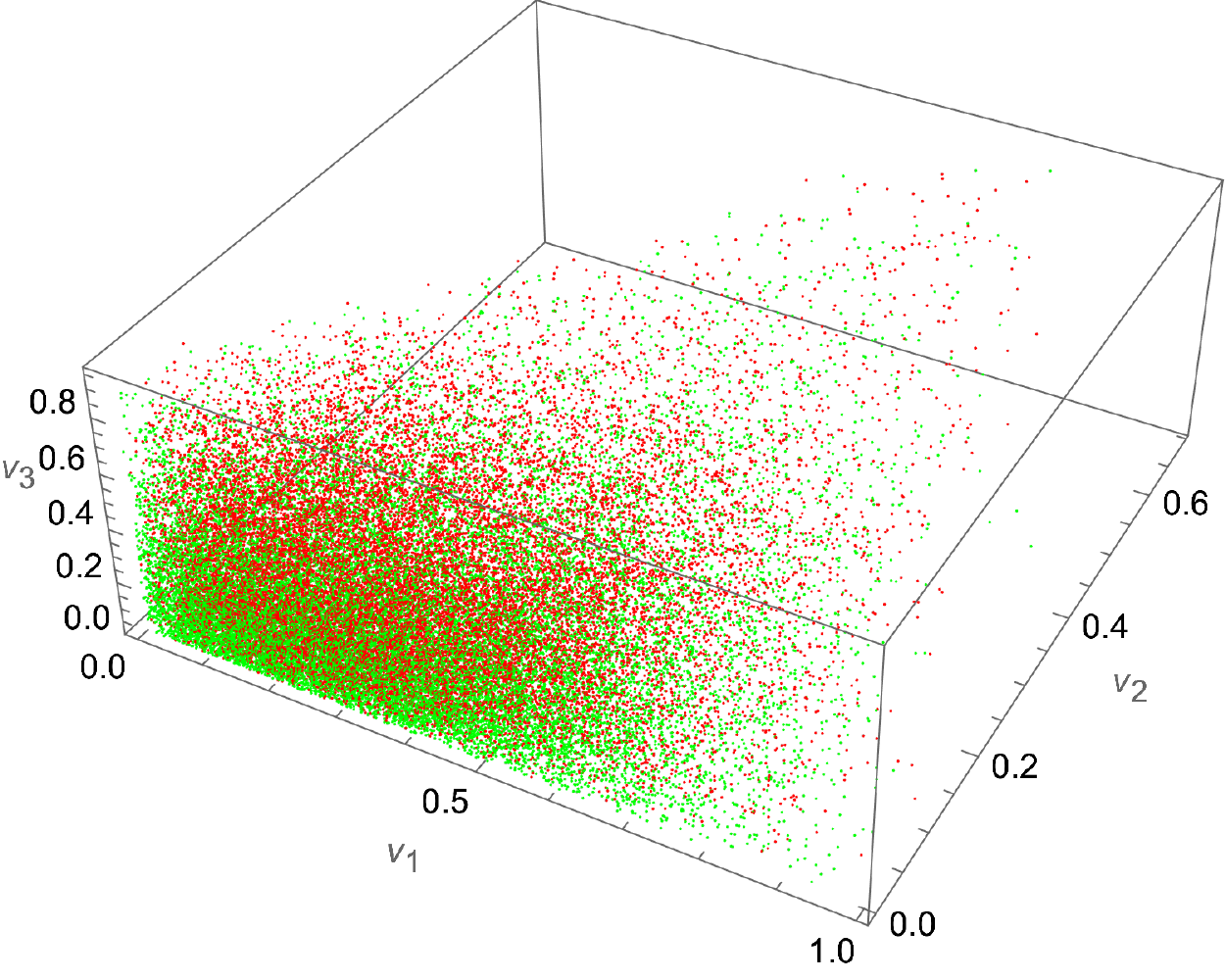}
    \caption{Rebit-retrit singular value ratios. There are twenty thousand red points corresponding to randomly selected separable states, and twenty thousand green points corresponding to randomly selected entangled states. Here, $v_1$ is the ratio of the second largest singular value of $V=D_2^{\frac{1}{2}} D_1^{-\frac{1}{2}}$ to the first, while $v_2$ is the ratio of the third  largest singular value to the first, and $v_3$ is the ratio of the smallest  singular value to the second largest. The mean location of the separable points is $\{0.336021, 0.0938245, 0.294796\}$ and that of the entangled points, $\{0.335971, 0.0717335, 0.225009\}$. 
}
    \label{fig:SVDrebitretrit}
\end{figure}

In Fig.~\ref{fig:SVDrebitretrit}, each selected density matrix is plotted according to its three singular value ratios. Here, $v_1$ is the ratio of the second largest singular value of $V$ to the first, while $v_2$ is the ratio of the third  largest singular value to the first, and $v_3$ is the ratio of the smallest  singular value to the second largest.

The correlation matrix for the three singular-value ratios $v_1,v_2,v_3$ based on the 
sampled twenty thousand  {\it separable} rebit-retrit density matrices (represented by red points) (Fig.~\ref{fig:SVDrebitretrit}) is 
\begin{equation}
 \left(
\begin{array}{ccc}
 1. & 0.554743 & -0.154844 \\
 0.554743 & 1. & 0.624914 \\
 -0.154844 & 0.624914 & 1. \\
\end{array}
\right) .  
\end{equation}
We note that $v_1$ and $v_3$ are negatively correlated.

The correlation matrix for the three singular-value ratios $v_1,v_2,v_3$ based on the sampled twenty thousand {\it entangled} rebit-retrit density matrices (represented by green points) (Fig.~\ref{fig:SVDrebitretrit}) is somewhat similar in nature
\begin{equation}
\left(
\begin{array}{ccc}
 1. & 0.507305 & -0.109214 \\
 0.507305 & 1. & 0.667493 \\
 -0.109214 & 0.667493 & 1. \\
\end{array}
\right)
\end{equation}

\subsubsection{Qubit-qutrit case}
\begin{figure}
    \centering
    \includegraphics{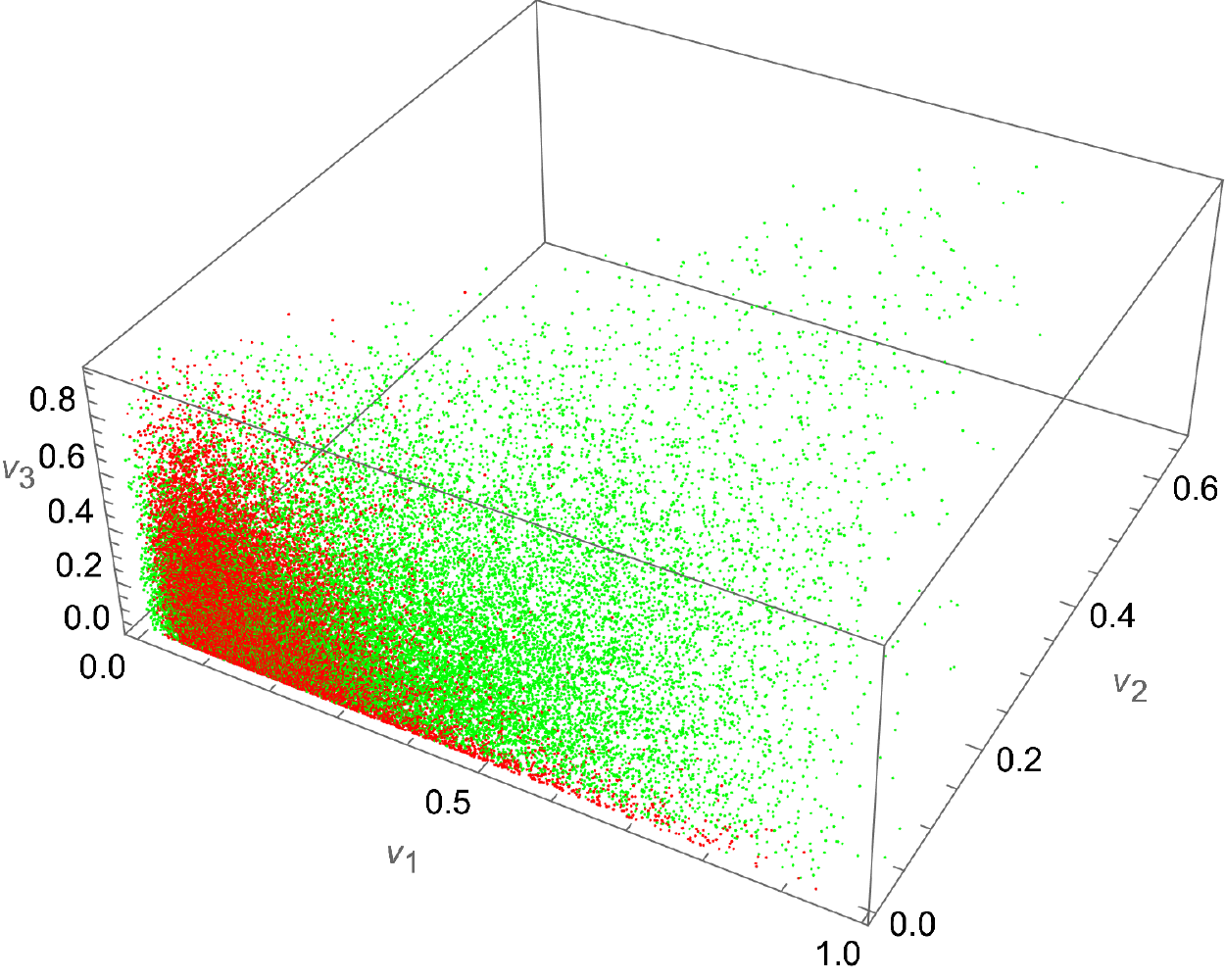}
    \caption{Qubit-qutrit singular value ratios. There are twenty thousand red points corresponding to randomly selected separable states, and twenty thousand green points corresponding to randomly selected entangled states. Here, $v_1$ is the ratio of the second largest singular value of $V= D_2^{\frac{1}{2}} D_1^{-\frac{1}{2}}$ to the first, while $v_2$ is the ratio of the third  largest singular value to the first, and $v_3$ is the ratio of the smallest  singular value to the second largest. The mean location of the separable points is $\{.164521, 0.0209839, 0.173539\}$ and that of the entangled points, $\{0.335391, 0.072119, 0.226911\}$. 
    }
    \label{fig:SVDqubitqutrit}
\end{figure}

In Fig.~\ref{fig:SVDqubitqutrit}, each selected density matrix is similarly plotted according to its three singular value ratios.

The correlation matrix for the three singular-value ratios $v_1,v_2,v_3$ based on the sampled twenty thousand {\it entangled} qubit-qutrit density matrices (represented by green points) (Fig.~\ref{fig:SVDqubitqutrit}) is 
\begin{equation}
\left(
\begin{array}{ccc}
 1. & 0.50146 & -0.111931 \\
 0.50146 & 1. & 0.66744 \\
 -0.111931 & 0.66744 & 1. \\
\end{array}
\right).
\end{equation}

The correlation matrix for the three singular-value ratios $v_1,v_2,v_3$ based on the 
sampled twenty thousand  {\it separable} qubit-qutrit density matrices (represented by red points) (Fig.~\ref{fig:SVDqubitqutrit}) is 
\begin{equation}
\left(
\begin{array}{ccc}
 1. & 0.0707637 & -0.418406 \\
 0.0707637 & 1. & 0.663828 \\
 -0.418406 & 0.663828 & 1. \\
\end{array}
\right)   
\end{equation}
is somewhat different in nature than the previous three correlation matrices.

\subsection{Joint distribution when $D_1$ and $D_2$ are themselves diagonal}
\begin{figure}
    \centering
    \includegraphics{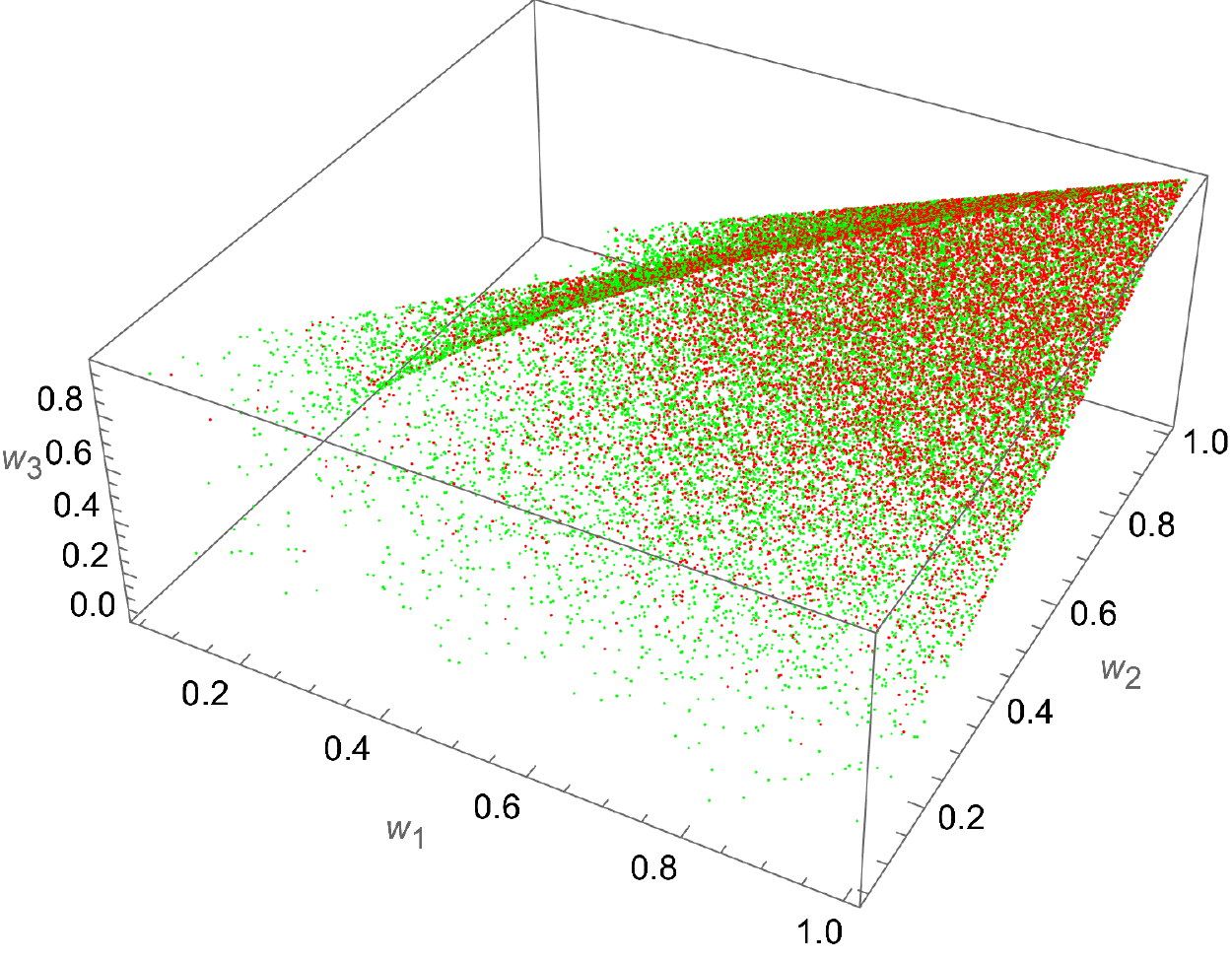}
    \caption{Rebit-retrit singular value ratios ($w_1, w_2, w_3$) based on {\it diagonal} $3 \times 3$ $D_1, D_2$. There are twenty thousand red points corresponding to randomly selected separable states, and twenty thousand green points corresponding to randomly selected entangled states.}
    \label{fig:Diagonalrebitretrit}
\end{figure}

Now, in \cite{slater2017master}, it was found that if the two $2 \times 2$ diagonal blocks $D_1$ and $D_2$ were  diagonal matrices, the Lovas-Andai variable $\sigma (V)$ coincided with 
the variable $\sqrt{\frac{\rho_{11} \rho_{44}}{\rho_{22} \rho_{33}}}$--or its reciprocal--that had been used by Slater in a number of separability probability studies.
In the $6 \times 6$ setting if the two $2 \times 2$ diagonal blocks $D_1$ and $D_2$ are  diagonal, the singular-value ratios are
\begin{equation}
w_=\sqrt{\frac{\rho_{22} \rho_{44}}{\rho_{11} \rho_{55}}}, w_1=\sqrt{\frac{\rho_{33} \rho_{44}}{\rho_{11} \rho_{66}}}, w_1=\sqrt{\frac{\rho_{33} \rho_{55}}{\rho_{22} \rho_{66}}} \end{equation}
or their reciprocals. In Fig.~\ref{fig:Diagonalrebitretrit}--as in the preceding two figures--we show these values (or their reciprocals) for twenty thousand red points corresponding to randomly selected separable states, and twenty thousand green points corresponding to randomly selected entangled states.

In the $6 \times 6$ {\it correlation} matrix for the six twenty-thousand long vectors of singular value ratios for the rebit-retrit separable and entangled samples based on {\it diagonal} $3 \times 3$ $D_1, D_2$ (Fig.~\ref{fig:Diagonalrebitretrit}), the off-diagonal entries lie in [-0.0108477,0.223996], indicative of rather weak correlations between the six vectors.

\begin{figure}
    \centering
    \includegraphics{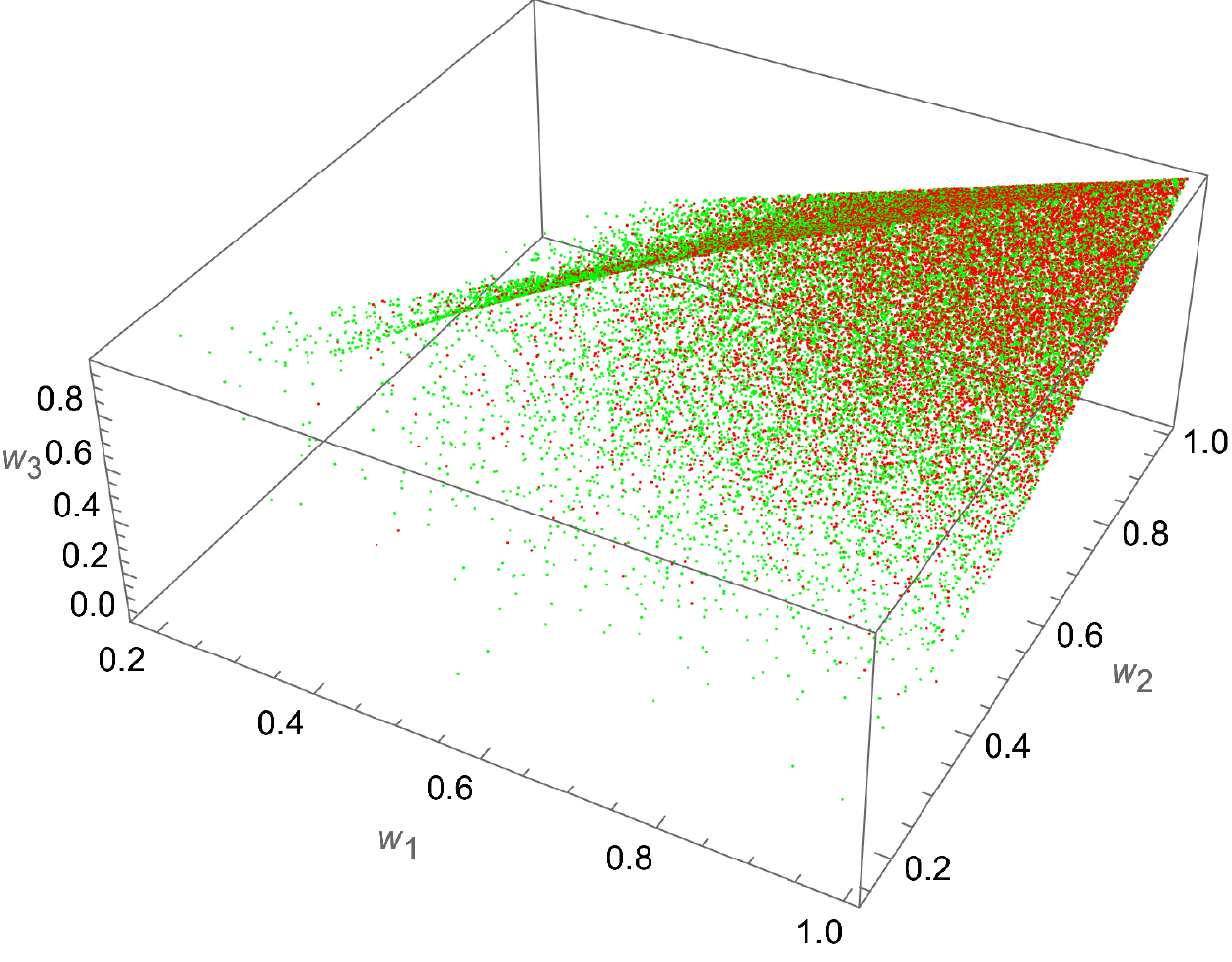}
    \caption{Qubit-qutrit singular value ratios $w_1, w_2, w_3$) based on {\it diagonal} $3 \times 3$ $D_1, D_2$s. There are twenty thousand red points corresponding to randomly selected separable states, and twenty thousand green points corresponding to randomly selected entangled states.}
    \label{fig:Diagonalqubitqutrit}
\end{figure}

\begin{figure}
    \centering
    \includegraphics{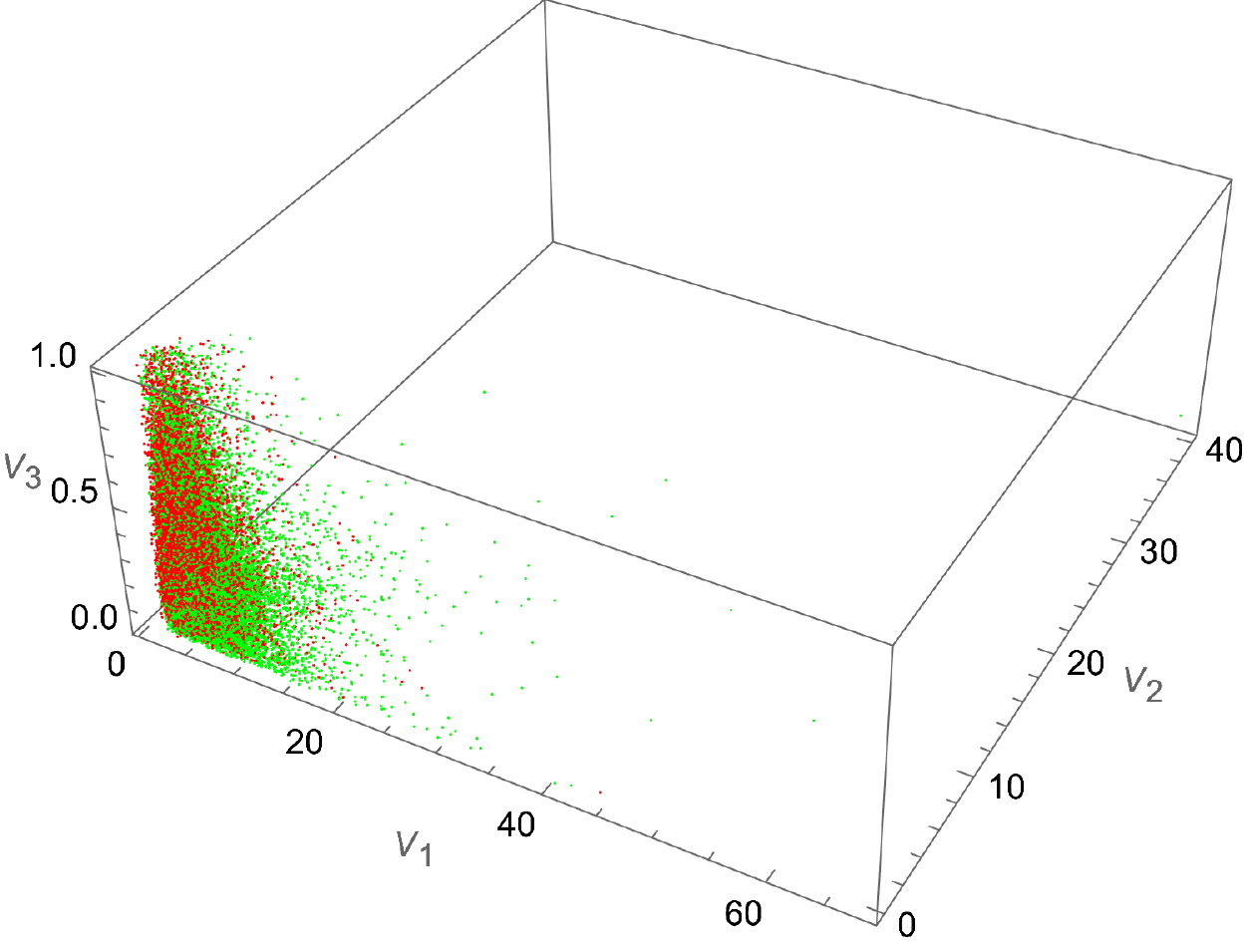}
    \caption{Rebit-retrit singular values themselves (not ratios). There are twenty thousand red points corresponding to randomly selected separable states, and twenty thousand green points corresponding to randomly selected entangled states. Here, $V_1 \geq V_2 \geq V_3$ are the singular values. The mean location of the separable points is $\{3.96817, 1.19803, 0.311059\}$ and that of the entangled points, $\{4.97101, 1.47055, 0.267268\}$. 
}
    \label{fig:ABSrebitretrit}
\end{figure}

\begin{figure}
    \centering
    \includegraphics{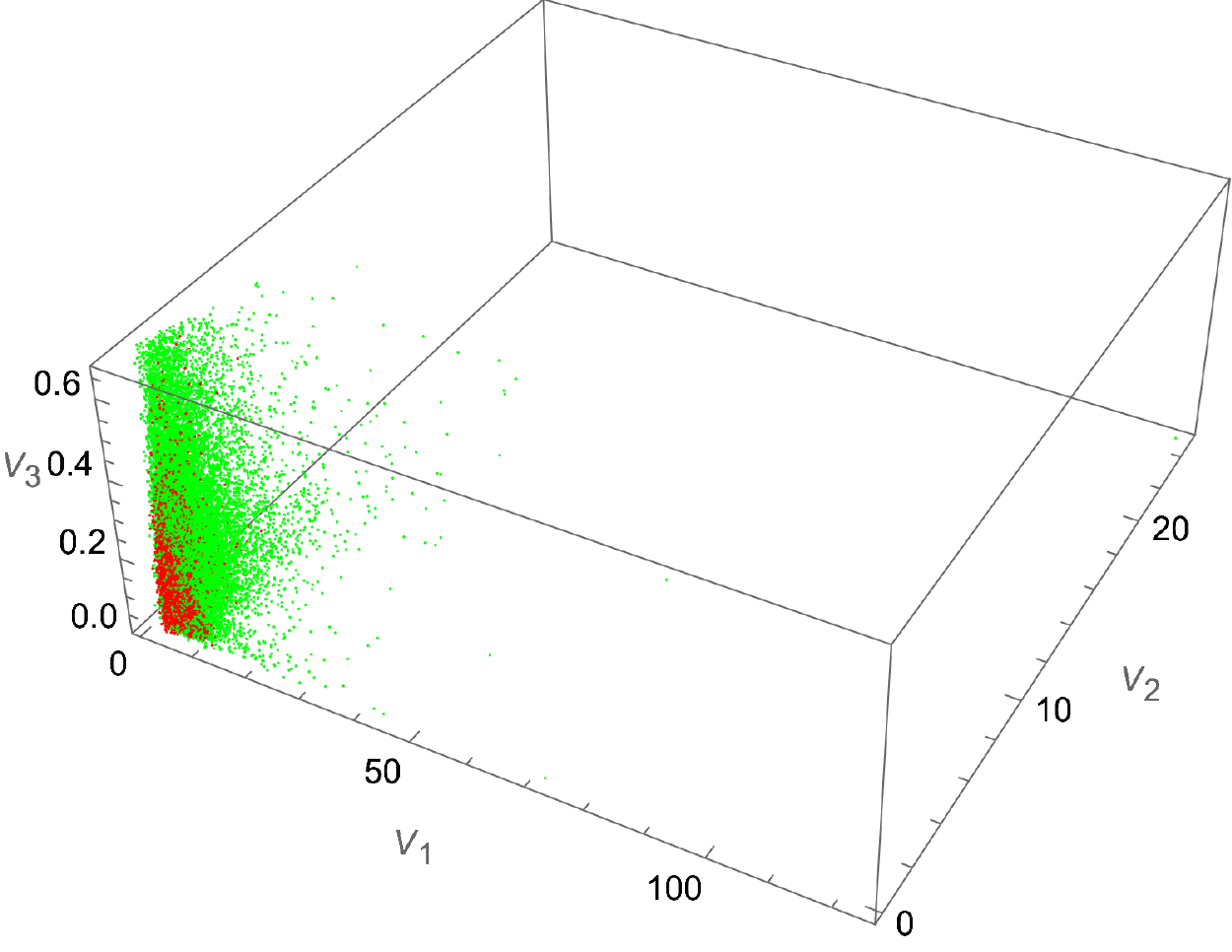}
    \caption{Qubit-qutrit singular values themselves (not ratios). There are twenty thousand red points corresponding to randomly selected separable states, and twenty thousand green points corresponding to randomly selected entangled states. Here, $V_1 \geq V_2 \geq V_3$ are the singular values. The mean location of the separable points is $\{4.12892, 0.615074, 0.0807939\}$ and that of the entangled points, $\{4.95325, 1.47314, 0.269761\}$. 
}
    \label{fig:ABSqubitqutrit}
\end{figure}

In the $6 \times 6$ {\it correlation} matrix for the six twenty-thousand long vectors of singular value ratios for the qubit-qutrit separable and entangled samples based on {\it diagonal} $3 \times 3$ $D_1, D_2$ (Fig.~\ref{fig:Diagonalqubitqutrit}), the off-diagonal entries lie in [-0.00812645,0.229486], again indicative of rather weak correlations between the six vectors.
Now, let us repeat the forms of analysis employed to yield Figs.~\ref{fig:SVDrebitretrit} and \ref{fig:SVDqubitqutrit}, but now use as the three axes, the singular values themselves, rather than ratios of them (Figs~\ref{fig:ABSrebitretrit},\ref{fig:ABSqubitqutrit}).

\subsection{Separability probabilities as functions of the 
singular value ratios}
\subsubsection{Rebit-retrit case}
\begin{figure}
    \centering
    \includegraphics{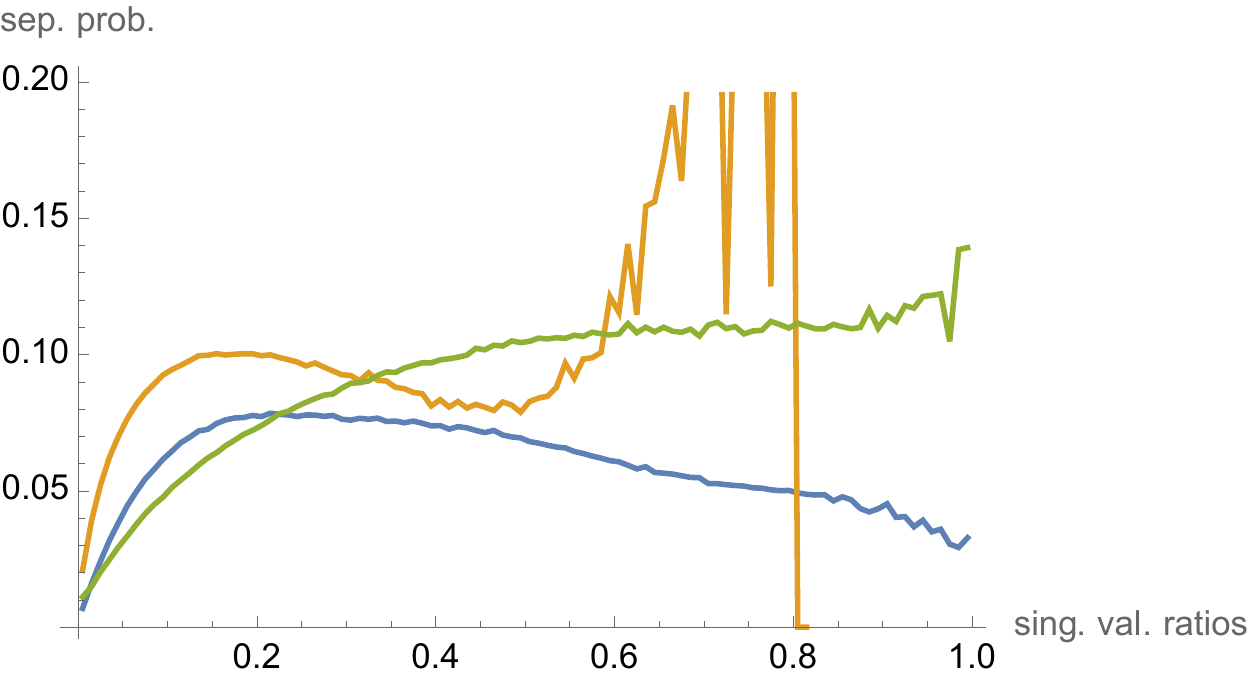}
    \caption{The separability probability, based upon thirty-five million random rebit-retrit density matrix realizations, as a function of the 
singular value ratios. The blue--lowest-value at 1--curve is based on the ratio of the second largest singular value to the first, while the steeply uprising yellow curve $v_2$ is based on the ratio of the third  largest singular value to the first, and the green curve is based on the ratio of the smallest  singular value to the second largest.}
    \label{fig:RebitRetritsepprob}
\end{figure}

Changing our focus, in Fig.~\ref{fig:RebitRetritsepprob}, based upon thirty-five million random rebit-retrit density matrix realizations, we show the separability probability as a function of 
the singular value ratios. The blue--lowest-value at 1--curve is based on the ratio of the second largest singular value to the first, while the steeply uprising yellow curve  is based on the ratio of the third  largest singular value to the first, and the green curve is based on the ratio of the smallest  singular value to the second largest.

In the two-rebit instance successfully analyzed by Lovas and Andai, the separability probability function took the form \cite[eq. (1)]{lovas2017invariance}
\begin{equation} \label{BasicFormula}
\tilde{\chi}_1 (\varepsilon ) = 1-\frac{4}{\pi^2}\int\limits_\varepsilon^1 
\left(
s+\frac{1}{s}-
\frac{1}{2}\left(s-\frac{1}{s}\right)^2\log \left(\frac{1+s}{1-s}\right)
\right)\frac{1}{s}
\mbox{d}  s 
\end{equation}
\begin{displaymath}
 = \frac{4}{\pi^2}\int\limits_0^\varepsilon
\left(
s+\frac{1}{s}-
\frac{1}{2}\left(s-\frac{1}{s}\right)^2\log \left(\frac{1+s}{1-s}\right)
\right)\frac{1}{s}
\mbox{d} s.
\end{displaymath}
Further \cite[eq. (2)]{lovas2017invariance}, 
$\tilde{\chi}_1 (\varepsilon )=\tilde{\eta}_1 (\varepsilon )$ has a closed form,
\begin{equation} \label{poly}
\frac{2 \left(\varepsilon ^2 \left(4 \text{Li}_2(\varepsilon )-\text{Li}_2\left(\varepsilon
   ^2\right)\right)+\varepsilon ^4 \left(-\tanh ^{-1}(\varepsilon )\right)+\varepsilon ^3-\varepsilon
   +\tanh ^{-1}(\varepsilon )\right)}{\pi ^2 \varepsilon ^2},    
\end{equation}
where the polylogarithmic function is defined by the infinite sum
	\begin{equation*}
		\text{Li}_s (z) =
		\sum\limits_{k=1}^\infty 
		\frac{z^k}{k^s},
	\end{equation*}
for arbitrary complex $s$ and for all complex arguments $z$ with $|z|<1$. 

Lovas and Andai noted that it is was "somewhat interesting that the identity function approximates $\tilde{\chi}_1(\varepsilon)$ well". However, the the three curves in Fig.~\ref{fig:RebitRetritsepprob} appear quite nonlinear in nature.
\subsubsection{Qubit-qutrit case}

\begin{figure}
    \centering
    \includegraphics{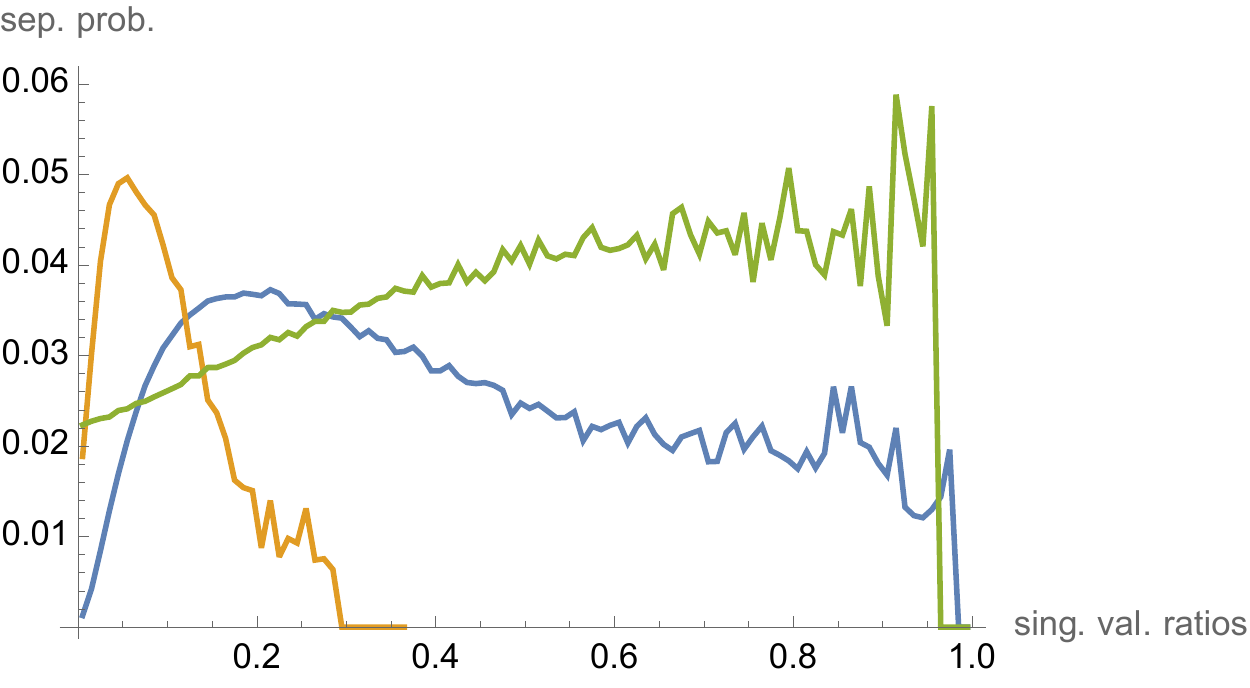}
    \caption{The separability probability, based upon twenty million random qubit-qutrit density matrix realizations as a function of the 
singular value ratios. The blue curve is based on the ratio of the second largest singular value to the first, while the  yellow, steeply declining curve is based on the ratio of the third  largest singular value to the first, and the green, generally ascending curve is based on the ratio of the smallest  singular value to the second largest.}
    \label{fig:QubitQutritSepProb}
\end{figure}

\begin{figure}
\includegraphics{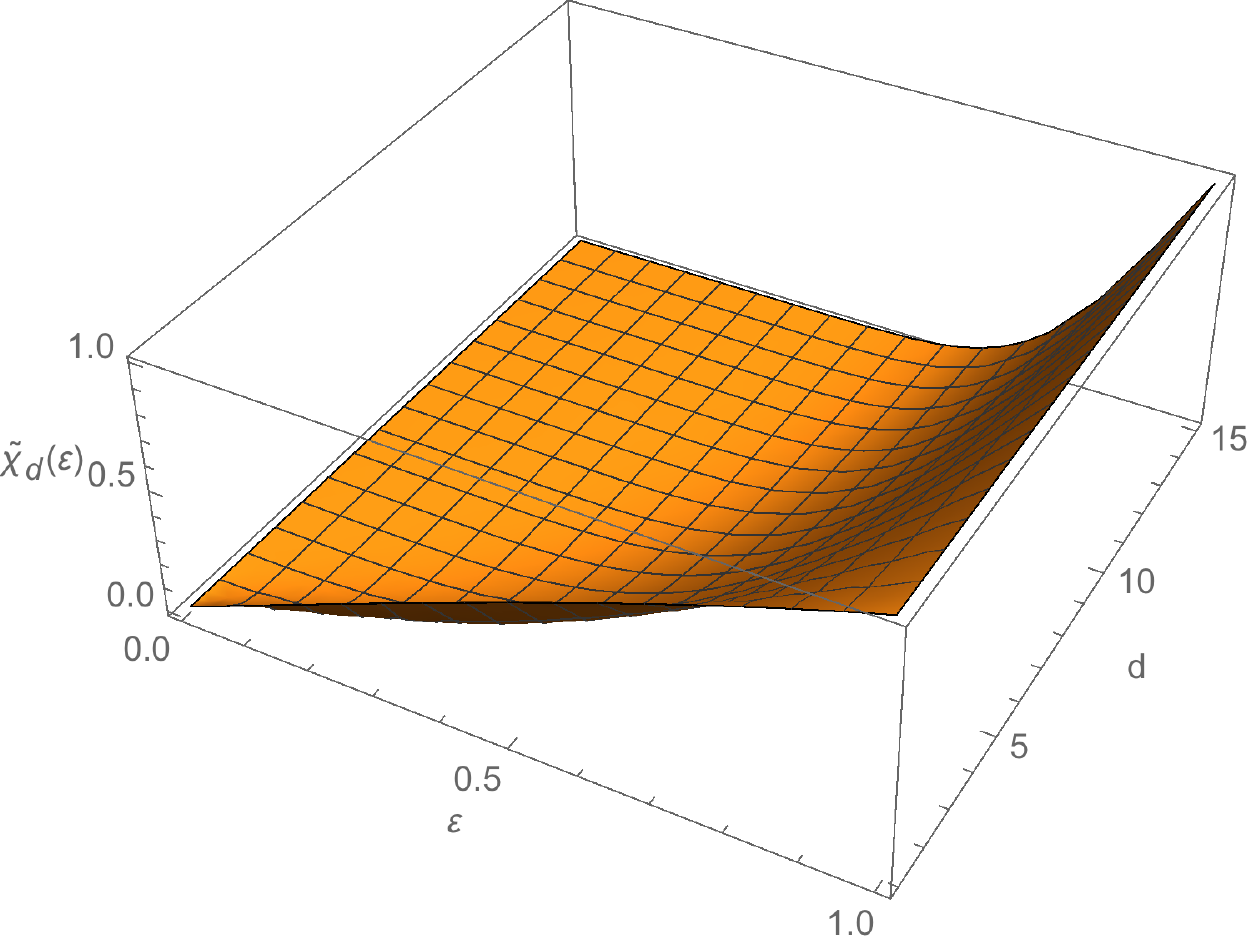}
\caption{\label{fig:MasterFormula}Lovas-Andai master formula (\ref{Ourformula2}) for $\tilde{\chi_d} (\varepsilon )$. The value $d=1$ yields the original Lovas-Andai two-rebit formula (\ref{BasicFormula}) that established the conjectured separability probability of $\frac{29}{64}$, the value $d=2$ (\ref{myformula})  leads to $\frac{8}{33}$ as the two-qubit Hilbert-Schmidt separability probability, and $d=4$, gives $\frac{26}{323}=\frac{2 \cdot 13}{17 \cdot 19} \approx 0.0804954$ in the two-quater[nionic]-bit instance.}
\end{figure}

In Fig.~\ref{fig:QubitQutritSepProb}, based upon twenty million random qubit-qutrit density matrix realizations, we show the separability probability as a function of the 
singular value ratios. The blue curve is based on the ratio of the second largest singular value to the first, while the  yellow steeply declining curve is based on the ratio of the third  largest singular value to the first, and the green generally ascending curve is based on the ratio of the smallest  singular value to the second largest.
In the two-qubit instance, the separability probability function takes the form \cite[eq. (42)]{slater2017master}
\begin{equation} \label{myformula}
\tilde{\chi}_2(\varepsilon)  = \frac{1}{3} \varepsilon^2 (4 -\varepsilon^2). 
\end{equation}
More generally still \cite[eq. (70), Fig. 22]{slater2017master} (Fig.~\ref{fig:MasterFormula} in this paper),
\begin{equation} \label{Ourformula2}
\tilde{\chi_d} (\varepsilon )=
\end{equation}
\begin{displaymath}
\frac{\varepsilon ^d \Gamma (d+1)^3 \,
   _3\tilde{F}_2\left(-\frac{d}{2},\frac{d}{2},d;\frac{d}{2}+1,\frac{3
   d}{2}+1;\varepsilon ^2\right)}{\Gamma \left(\frac{d}{2}+1\right)^2}
(the tilde indicating regularization of the hypergeometric function).
\end{displaymath}

\subsection{$8 \times 8$ density matrices} \label{EightByEight}
\begin{figure}
    \centering
    \includegraphics{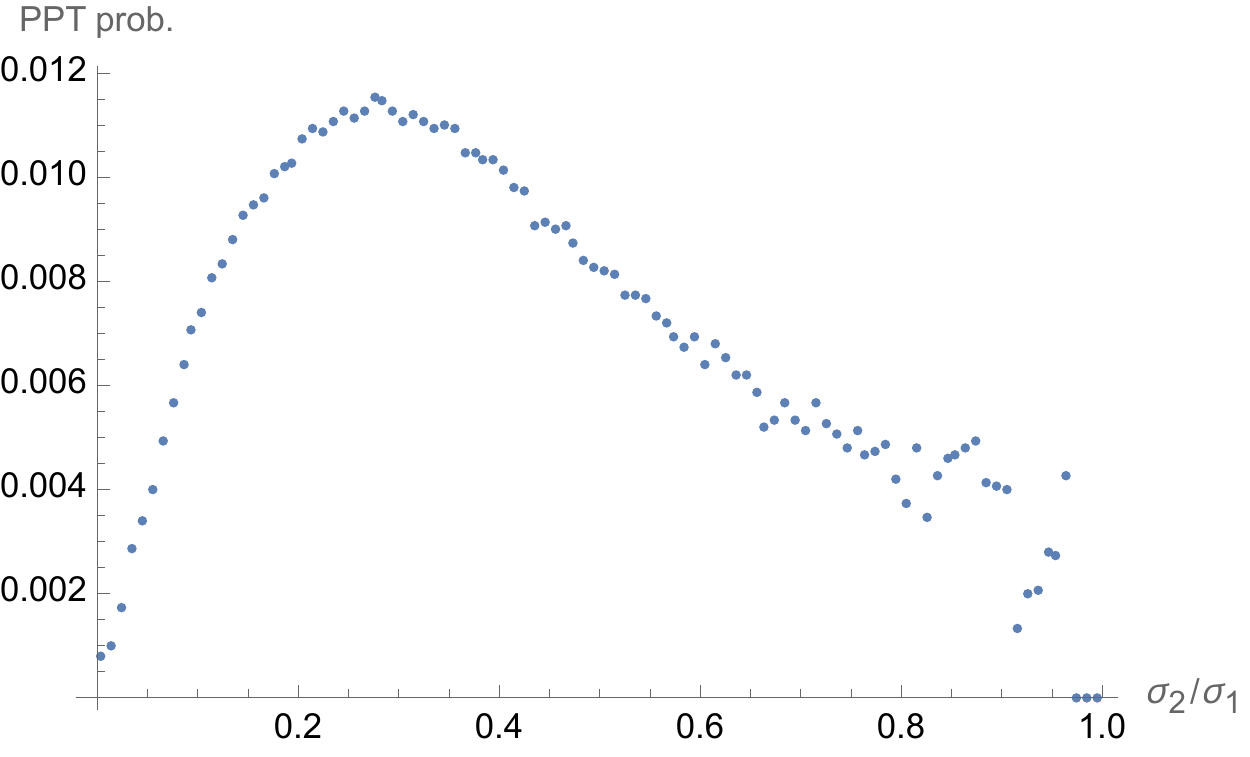}
    \caption{PPT probabilities as a function of the ratio of the second largest singular value ($\sigma_2$) of 
$V=D_2^{\frac{1}{2}} D_1^{-\frac{1}{2}}$ to the largest singular value ($\sigma_1$) of thirty-one million randomly generated--with respect to Hilbert-Schmidt measure \cite{al2010random}--$8 \times 8$ density matrices of the rebit-redit type.}
    \label{fig:RebitRedit}
\end{figure}

\begin{figure}
    \centering
    \includegraphics{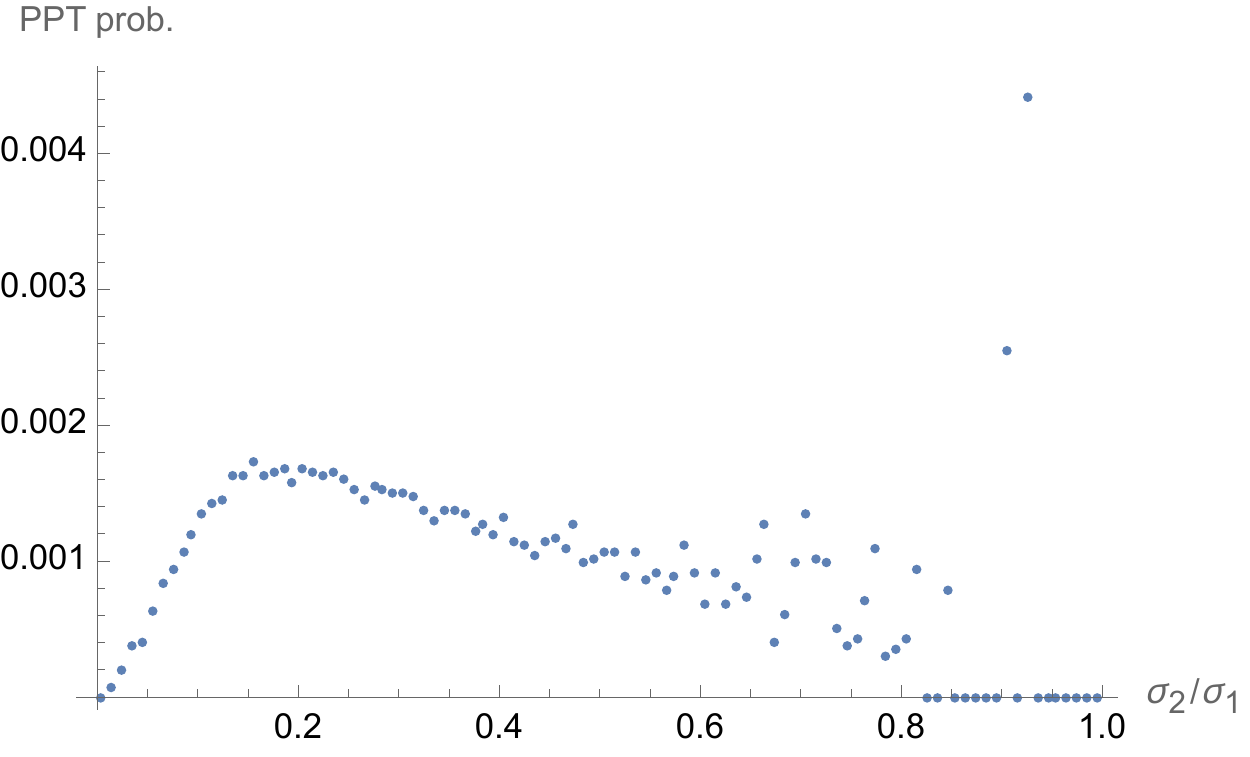}
    \caption{PPT probabilities as a function of the ratio of the second largest singular value ($\sigma_2$) of 
$V=D_2^{\frac{1}{2}} D_1^{-\frac{1}{2}}$ to the largest singular value ($\sigma_1$) of seventeen million randomly generated--with respect to Hilbert-Schmidt measure \cite{al2010random}--$8 \times 8$ density matrices of the qubit-qudit type.}
    \label{fig:QubitQudit}
\end{figure}

As an initial investigation into systems of higher dimensions than those of rebit-retrit and qubit-qutrits, we randomly generated--with respect to Hilbert-Schmidt measure \cite{al2010random}--$8 \times 8$ density matrices of the rebit-redit and qubit-qudit types.
In Figs.~\ref{fig:RebitRedit} and \ref{fig:QubitQudit}, we plot the PPT probabilities recorded as a function of the ratio of the second largest singular value ($\sigma_2$) of 
$V=D_2^{\frac{1}{2}} D_1^{-\frac{1}{2}}$ to the largest singular value ($\sigma_1$). (Conjectures of $\frac{201}{8192}= \frac{3 \cdot 67}{2^{13}} \approx 0.0245361$ and $\frac{16}{12375}= \frac{4^2}{3^2 \cdot 5^3 \cdot 11} \approx 0.001292929$ have been advanced for the PPT probabilities in these two cases \cite[sec. 3.5]{slater2018extensions} \cite{slater2019numerical}.) Now, $D_1, D_2$ are the $4 \times 4$ diagonal blocks of the density matrices.
\section{Counterparts of Lovas-Andai "separability function" $\tilde{\chi}_d (\varepsilon)$ in higher dimensions}
Of course, the challenge remains of incorporating rebit-redit and qubit-qudit singular value ratios into a higher-dimensional analogue of the Lovas-Andai "separability function" $\tilde{\chi}_d (\varepsilon)$,  for which a general formula 
\begin{equation} \label{Ourformula}
\tilde{\chi_d} (\varepsilon )=
\end{equation}
\begin{displaymath}
\frac{\varepsilon ^d \Gamma (d+1)^3 \,
   _3\tilde{F}_2\left(-\frac{d}{2},\frac{d}{2},d;\frac{d}{2}+1,\frac{3
   d}{2}+1;\varepsilon ^2\right)}{\Gamma \left(\frac{d}{2}+1\right)^2}
\end{displaymath}
(the tilde indicating regularization of the hypergeometric function) was reported in \cite[(eq. (70)]{slater2017master}.

To compute the two-rebit, two-qubit, two-quater[nionic]bit separability probabilities, one takes the ratio of 
\begin{equation}
\int\limits_{-1}^1\int\limits_{-1}^x  \tilde{\chi}_{d} \left(
\left.\sqrt{\frac{1-x}{1+x}}\right/ \sqrt{\frac{1-y}{1+y}}	
	\right)(1-x^2)^{d}(1-y^2)^{d} (x-y)^{d} \mbox{d} y\mbox{d} x 
	\end{equation}
	to
\begin{equation}
\int\limits_{-1}^1\int\limits_{-1}^x  (1-x^2)^{d}(1-y^2)^{d}(x-y)^{d}  \mbox{d} y \mbox{d} x
\end{equation}	
for $d=1,2,4$.

The function $\tilde{\chi_d} (\varepsilon )$ (\ref{Ourformula}) remarkably solves the challenging problem posed by Lovas and Andai
in their Conclusion section  \cite[sec. 6]{lovas2017invariance} (and solved by them for $d=1$). They wrote: ``The structure of the unit ball in operator norm of $2\times 2$ matrices plays a 
  critical role in separability probability of qubit-qubit and rebit-rebit 
  quantum systems.
It is quite surprising that the space of $2\times 2$ real or complex matrices 
  seems simple, but to compute the volume of the set
\begin{equation} \label{2by2matrix}
\Big\{\begin{pmatrix}a & b\\ c& e\end{pmatrix} \Big\vert\ a, b, c, e \in \mathbb{K},
 \norm{\begin{pmatrix} a & b\\ c& e\end{pmatrix}} <1,\ \
 \norm{\begin{pmatrix} a & \varepsilon b\\ \frac{c}{\varepsilon}& e
\end{pmatrix}}  <1 \Big\}
\end{equation}
  for a given parameter $\varepsilon\in [0,1]$, which is the value of 
  the function $\tilde{\chi}_{d}(\varepsilon)$, is a very challenging problem.
The gist of our considerations is that the behavior of the function 
  $\tilde{\chi}_{d}(\varepsilon)$ determines the separability probabilities with respect
  to the Hilbert-Schmidt measure.'' 
  
  The operator norm $\norm{\cdot}$ is the largest singular value or Schatten-$\infty$ norm. Let us note 
  that Gl{\"o}ckner studied functions on the quaternionic unit ball \cite{glockner2001functions}. For related discussions see \cite{mathoverflow,mathoverflow0,EuclideanNorm}.

The volume of the unit ball of $n \times n$ matrices in the operator norm is \cite{EuclideanVolume}
\begin{equation}
n! \prod_{k\leq n} \frac{ \pi^k }{ ((k/2)! \binom{2k}{k})} . 
\end{equation}
In particular, one has:  $\frac{2}{3 \pi^2}$ for $n=2$; $\frac{8}{45 \pi^4}$ for $n=3$; and $\frac{4}{1575 \pi^8}$ for $n=4$

Let us inquire as to what is the natural extension--if any--of this two-dimensional question posed by Lovas and Andai to three dimensions? Perhaps it would be the trivariate problem
\begin{equation} \label{3by3matrix}
\Big\{\begin{pmatrix}a & b & c\\ d & e & f\\g & h & i\end{pmatrix} \Big\vert\ a, b, c, e. f, g, h, i \in \mathbb{K},
 \norm{\begin{pmatrix}a & b & c\\ d & e & f\\g & h & i\end{pmatrix} } <1,\ \
 \norm{\begin{pmatrix}a & b \varepsilon_1 & c \varepsilon_2 \\ \frac{d}{\varepsilon_1} & e & f \varepsilon_3\\\frac{g}{\varepsilon_2} & \frac{h}{\varepsilon_3} & i\end{pmatrix}} <1 \Big\},
\end{equation}
or possibly once again a univariate one, with $\varepsilon_1=\varepsilon_2=\varepsilon_3$.
Conceivably, it would be hoped, the research reported above might pertain, in some manner, to this clearly highly imposing problem.
\subsection{Excellent numeric approximation of original Lovas-Andai separability probability function}
\begin{figure}
    \centering
    \includegraphics{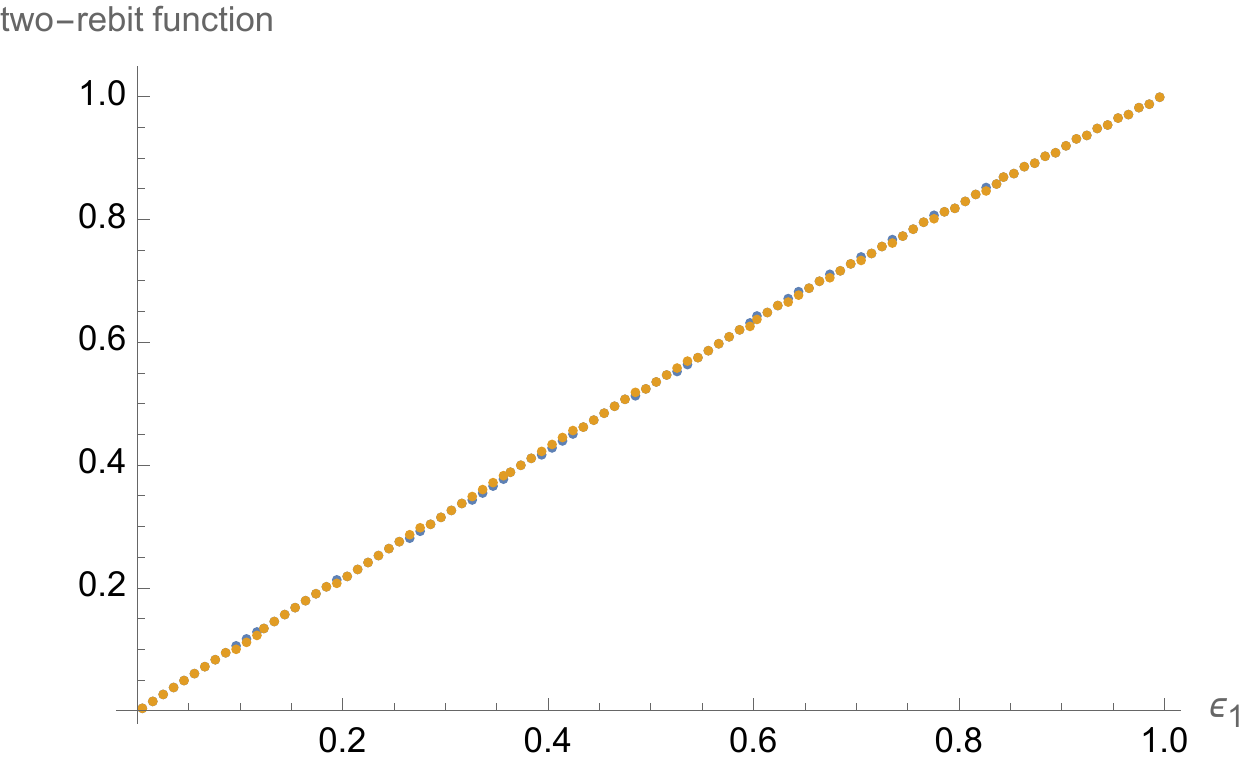}
    \caption{{\it Joint} plot of numerical approximation of Lovas-Andai two-rebit separability probability function $\tilde{\chi}_1 (\varepsilon )$ {\it along with} the essentially indistinguishable $\tilde{\chi}_1 (\varepsilon )$. The four entries of the $2 \times 2$ matrix (\ref{2by2matrix}) were randomly chosen from $[-10,10]$. Of five hundred million such quartets  generated, 20,634 yielded an operator norm less than 1, leading to their further analysis and the numerical approximation plot displayed.}.
    \label{fig:2by2matrix}
\end{figure}

Let us see, however, to what extent numerics can assist us in such a pursuit. To begin, as an exercise, let us demonstrate that we  can very well approximate the Lovas-Andai two-rebit separability function 
$\tilde{\chi}_1 (\varepsilon )$ (\ref{BasicFormula}) by 
numeric means. To such an end, for the entries $a, b, c, d$ of the leftmost $2  \times2$ matrix in (\ref{2by2matrix}), we randomly chose real numbers in $[-10,10]$. (We initially thought it would be ideal to use $[\infty,\infty]$--influenced, in part, by the non-imposition of limits by Lovas and Andai--but further examination of the literature and testing of ours indicated that the absolute values of all four entries--and, more generally, for higher-dimensional systems--could be no greater than 1.) Of five hundred million such choices, the number for which the corresponding operator norm was less than 1 was 20,634. For each such quartet, we tested whether the rightmost matrix in (\ref{2by2matrix}) also had operator norm less than 1 as a function of $\varepsilon$. Fig.~\ref{fig:2by2matrix} shows the resultant plot, {\it along with} the essentially indistinguishable $\tilde{\chi}_1 (\varepsilon )$,  close to $\varepsilon$ itself. Lovas and Andai in their Figure 1 noted that  $\tilde{\chi}_1 (\varepsilon)$ is interestingly close to $\varepsilon$ itself. 
\subsection{Rebit-retrit problem}
\begin{figure}
    \centering
    \includegraphics{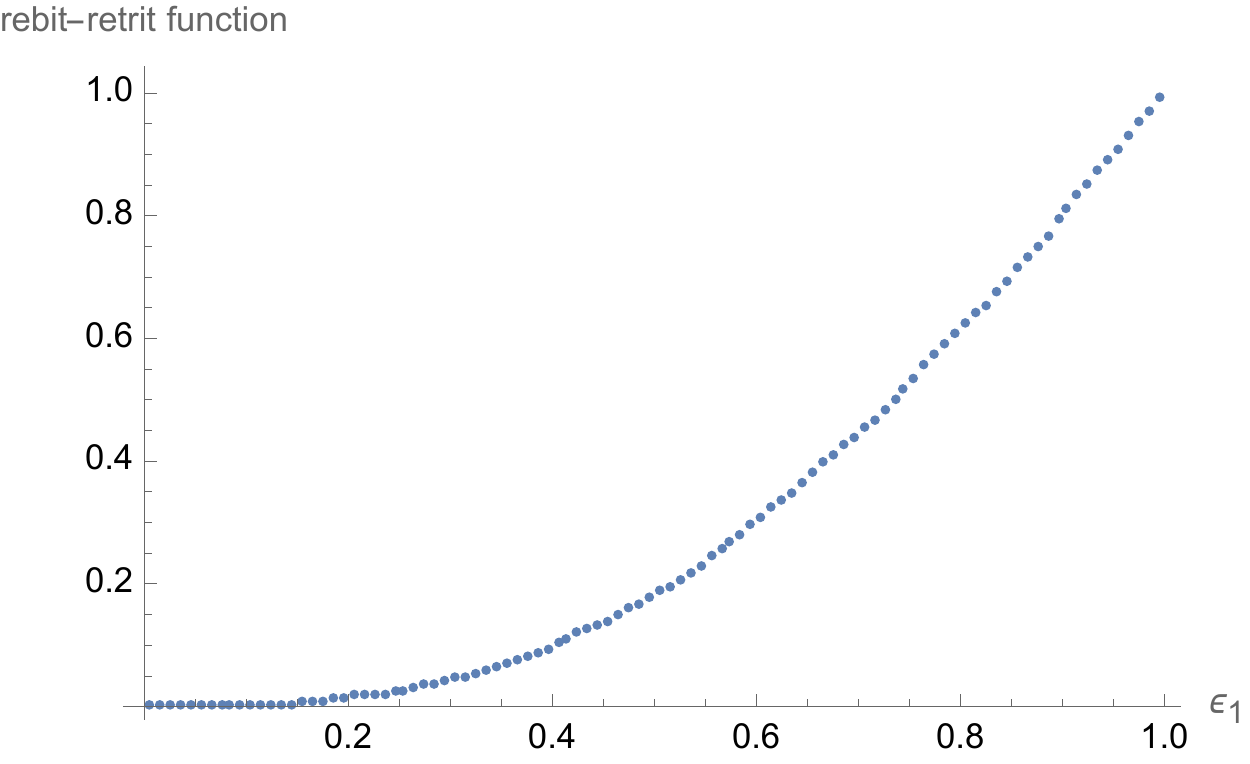}
    \caption{Numerical approximation of candidate Lovas-Andai rebit-retrit analogue of two-rebit  separability probability function $\tilde{\chi}_1 (\varepsilon )$. Of fifty million randomly generated $3 \times 3$ matrices with entries in [-2,2], 3,315 had operator norm less than 1, leading to their further analysis and the numerical approximation plot displayed.}
    \label{fig:3by3matrix}
\end{figure}

For the univariate version of the yet unexplored $3 \times 3$ problem just posed, we randomly chose the nine (real) entries of the leftmost matrix in (\ref{3by3matrix}) to lie in $[-2,2]$. Of fifty million generated, the number for which the corresponding operator norm was less than 1, was 3,315. For each such set of nine, we tested whether the rightmost matrix  (with $\varepsilon_1=\varepsilon_2=\varepsilon_3$), that is,  
\begin{equation}
\norm{\begin{pmatrix}a & b \varepsilon_1 & c \varepsilon_1 \\ \frac{d}{\varepsilon_1} & e & f \varepsilon_1\\\frac{g}{\varepsilon_1} & \frac{h}{\varepsilon_1} & i\end{pmatrix}}    
\end{equation}
also had operator norm less than 1 as a function of $\varepsilon_1$. Fig.~\ref{fig:3by3matrix} shows the resultant plot.
\subsection{Qubit-qutrit problem}
\begin{figure}
    \centering
    \includegraphics{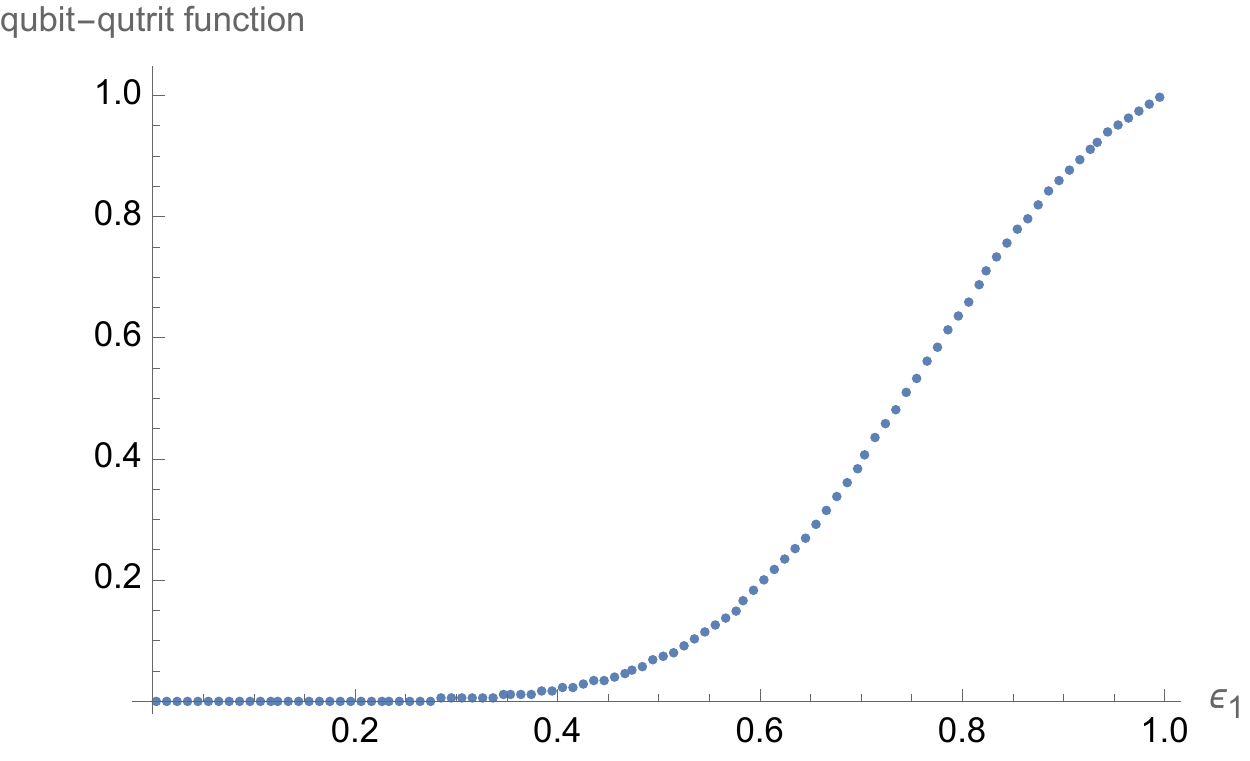}
    \caption{Numerical approximation of candidate Lovas-Andai qubit-qutrit analogue of two-qubit  separability probability function $\tilde{\chi}_2 (\varepsilon )$. Of 3,300,000 randomly generated $3 \times 3$ matrices with real and imaginary entries in $[-\frac{1}{2},\frac{1}{2}]$, 1,491,821 had operator norm less than 1, leading to their further analysis and the numerical approximation plot displayed.}
    \label{fig:3by3matrixQQ}
\end{figure}

Now, we allowed the nine entries of the $3 \times 3$ matrix to be complex in nature, rather than simply real. We randomly chose the real and imaginary components of the entries to lie in $[-\frac{1}{2},\frac{1}{2}]$. Of some 3,300,000 such choices, 1,491,821 yielded operator norm less than 1, leadng to their further analysis, resulting in Fig.~\ref{fig:3by3matrixQQ}.
\subsection{Two-retrit and two-qutrit problems}
\begin{figure}
    \centering
    \includegraphics{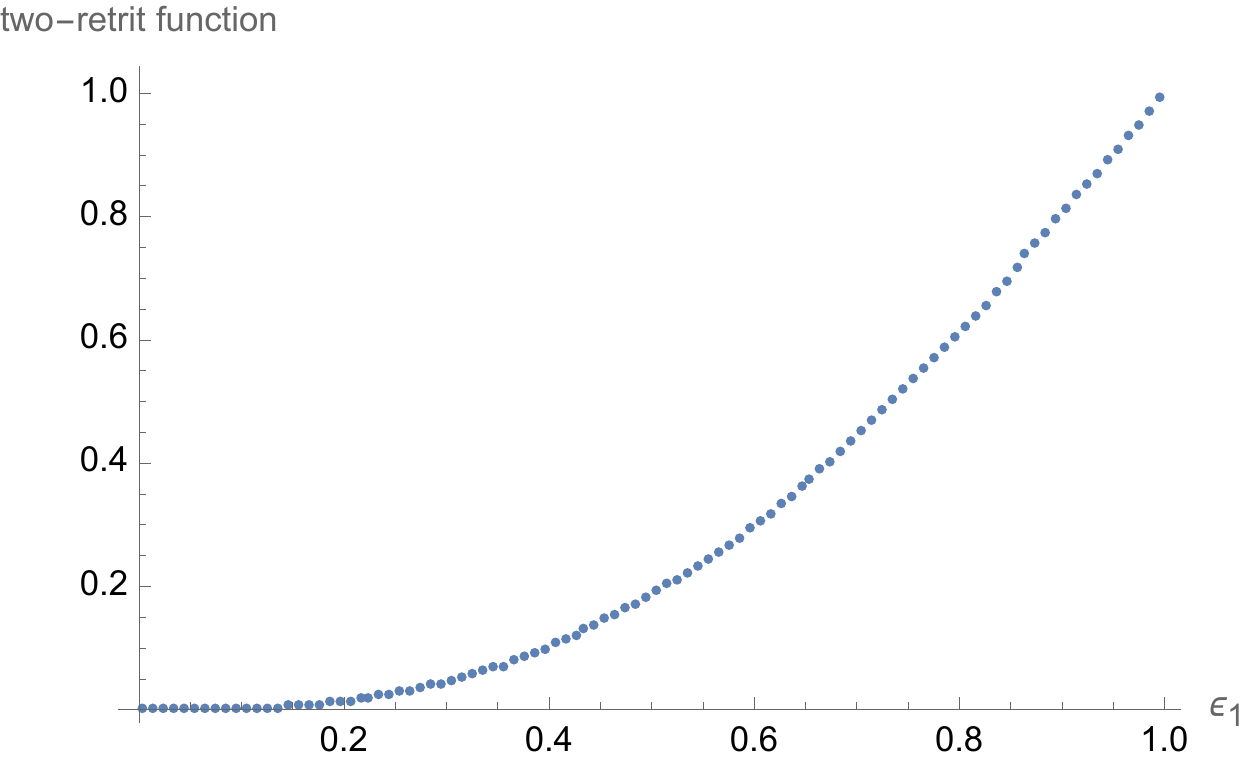}
    \caption{Numerical approximation of candidate Lovas-Andai two-retrit analogue of two-rebit  PPT probability function $\tilde{\chi}_1 (\varepsilon )$. Of 340,000,000 randomly generated $9 \times 9$ matrices with real  entries in $[-1,1]$, of the $3 \times 3$ matrices $D=D_1+D_2+D_3$ formed from them, 22,586 had operator norm less than 1, leading to their further analysis and the numerical approximation plot displayed.}
    \label{fig:3by3matrixTworetrits}
\end{figure}

\begin{figure}
    \centering
    \includegraphics{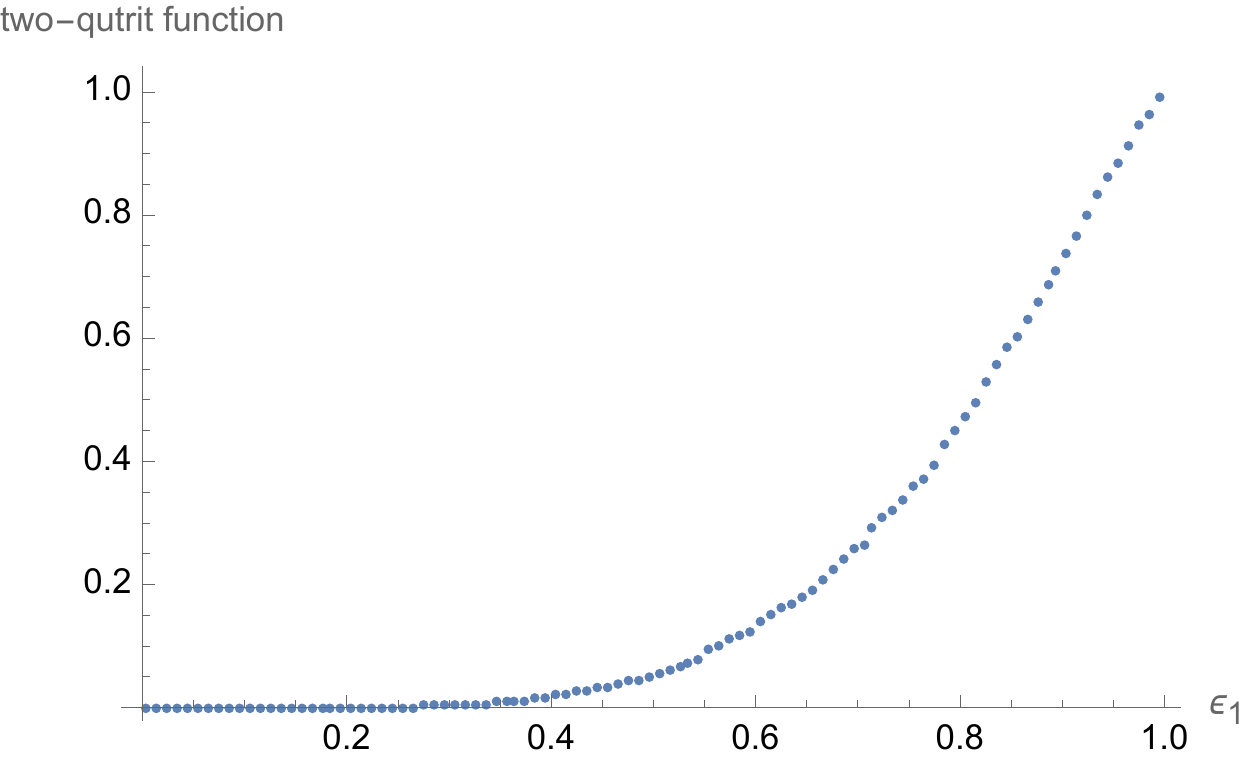}
    \caption{Numerical approximation of candidate Lovas-Andai two-qutrit analogue of two-qubit  PPT probability function $\tilde{\chi}_2 (\varepsilon )$.  Of 140,000,000 randomly generated $9 \times 9$ matrices with real  and imaginary entries entries in $[-\frac{1}{2},\frac{1}{2}]$, of the $3 \times 3$ matrices $D=D_1+D_2+D_3$ formed from them, 1,625 had operator norm less than 1, leading to their further analysis and the numerical approximation plot displayed. This curve is dominated by that of }
    \label{fig:3by3matrixTwoqutrits}
\end{figure}

For two-{\it retrit} and two-{\it qutrit} systems, there are clearly {\it three} natural diagonal blocks. 
So, is there an appropriate adaptation of the Lovas-Andai approach--involving the matrix  $V=D_2^{\frac{1}{2}} D_1^{-\frac{1}{2}}$--to that case? 
Let us, in this context, study  $D=D_1+D_2+D_3$, where the three $3 \times 3$ diagonal blocks of the $9 \times 9$ density matrices are indicated. In Fig.~\ref{fig:3by3matrixTworetrits}, we display the results of a two-retrit analysis, 
and in Fig.~\ref{fig:3by3matrixTwoqutrits}, that of a two-qutrit analysis.  A joint plot of these last two curves shows the two-retrit one fully dominating the two-qutrit one.

It would be of interest to examine the case of $8 \times 8$ density matrices (sec.~\ref{EightByEight}), using not two $4 \times 4$ diagonal blocks, but rather the four $2 \times 2$ diagonal blocks.

\section{Concluding remarks}
Let us note, in closing, that the doctoral dissertation of A. Lovas has several mentions of qubit-qutrits \cite{AttilaPhD}, particularly sec. 3.3. However, since it is written in Hungarian, it is rather
challenging to fully grasp the assertions made within it. 

In a personal communication, Lovas wrote "Chapter 3 deals with separability problems in composite quantum systems that can be described by an 2n-dimensional Hilbert-space. Especially, we have generalized the conjecture of Milz and Strunz (Section 3.3). A new parameterization was introduced (See Equations 3.29-3.31) that is more convenient to handle separability. 

We have proved that the $D_{2n}$ state space is diffeomorphic to the product of the $D_{n}$ state space, the [-I,I] operator interval and the unit sphere of $n \times n$ matrices with respect to the operator norm (Schatten-$\infty$ norm).

Theorem 3.3.3. characterizes the set of PPT states in this new parameterization. Theorem 3.3.5 tells us how come separability functions into play and how look like their higher dimensional generalizations."

\begin{acknowledgements}
This research was supported by the National Science Foundation under Grant No. NSF PHY-1748958.
\end{acknowledgements}

\bibliography{main}

\end{document}